\documentclass{spie}

\usepackage{graphicx}

\title{Fluctuation-induced first order transition due to Griffiths anomalies of the Cluster glass phase}

\author{Matthew J. Case\supit{a} and V. Dobrosavljevi\'c\supit{a,b}
\skiplinehalf
\supit{a}National High Magnetic Field Laboratory, Tallahassee, Florida, USA;\\
\supit{b}Florida State University, Tallahassee, Florida, USA
}

\authorinfo{Further author information: (Send correspondence to V. D.)\\
V. D.: E-mail: vlad@magnet.fsu.edu, Telephone: +1 850 644 5693\\
M. J. C.: E-mail: case@magnet.fsu.edu, Telephone: +1 850 644 1512
}

\begin{document}
\maketitle

\begin{abstract}
In itinerant magnetic systems with disorder, the quantum Griffiths phase at
$T=0$ is unstable to formation of a cluster glass (CG) of frozen droplet
degrees of
freedom.  In the absence of the fluctuations associated with these degrees of
freedom, the transition from the paramagnetic Fermi liquid (PMFL) to the
ordered phase proceeds via a conventional second-order quantum
phase transition.  However, when the Griffiths anomalies due to the broad
distribution of local energy scales are included, the transition is driven
first-order via a novel mechanism for a fluctuation induced first-order
transition.  At higher temperatures, thermal effects restore the
transition to second-order.  Implications of the enhanced non-Ohmic
dissipation in the CG phase are briefly discussed.
\end{abstract}


\section{INTRODUCTION}

When quenched disorder is introduced into a magnetic system in its ordered
phase, exponentially rare defect-free regions or {\it droplets} can form which
are nearly
magnetically ordered, even if the random magnet itself is not.  The system is
correlated over the length scale of the individual droplets and
long time scale (compared to that of a single spin) fluctuations are induced
due to coherent flipping of these large volumes of spins.  Such slow
fluctuations lead to singularities of the free energy in what is known as the
Griffiths phase.\cite{Griffiths:1969be}
Classically, these Griffiths singularities are weak, essential
singularities\cite{PhysRevB.10.4665}
and can only make exponentially small corrections to thermodynamic quantities.
However, near quantum phase transitions, disorder is perfectly correlated in
imaginary time making its effect much stronger than in the corresponding
classical problem.  Recently, much progress has been made in understanding the
effects of disorder on quantum critical points by focusing on the dynamics in
the quantum Griffiths phase
(QGP).\cite{Fisher:1992qf,fisher:6411,Motrunich:2000zt,PhysRevB.66.174433,
vojta:107202,vojta:045438,dobrosavljevic:187203}

One compelling reason for much the focus in this area is in understanding the
role QGP anomalies play in the non-Fermi liquid behavior of disordered strongly
correlated electron systems.\cite{0034-4885-68-10-R02}  For example, in the
random
transverse field Ising model in dimensions 1\cite{Fisher:1992qf}, 2 and
3,\cite{Motrunich:2000zt} it has been shown that a quantum Griffiths phase
characterized by a broad distribution of local energy scales precedes a $T=0$
transition in the universality class of the infinite-randomness fixed point
(IRFP).  The distribution of energies follows a power-law form with a
continuously varying exponent $\alpha=d/z'$ where $d$ is the dimensionality and
$z'$ is a dynamical critical exponent which diverges at the IRFP.  This
power-law distribution function leads to strong corrections to Fermi liquid
scaling in the QGP.  In particular, thermodynamic quantities become divergent
at $T=0$ showing that quantum Griffiths anomalies can have significant impact
on the behavior of random systems near their critical points.

Significant progress has been made recently by considering the rare regions in
the Griffiths phase as independent droplet degrees of freedom, and an elegant
classification scheme was proposed based only on general symmetry arguments
(for a review see Ref.~\citenum{0305-4470-39-22-R01}).  It was shown that
the effects of quantum Griffiths anomalies range from weak, classical
corrections, to complete destruction (rounding) of the phase transition.
However, these arguments apply only to insulating magnets for which the
droplets can truly be considered as independent.  By considering interactions
induced between droplets in a metallic host, it was
recently shown\cite{dobrosavljevic:187203} that the QGP is unstable to
formation of a cluster glass phase (CGP), and the conventional quantum
critical point is destroyed.  In the present work, we will examine the same
model as in Ref.~\citenum{dobrosavljevic:187203} and shown that fluctuations
due to Griffiths phase anomalies have the further effect to drive the
transition first-order.  At higher temperatures, thermal effects act to
suppress these fluctuations and a second-order transition is restored.

The paper is organized as follows: Sections \ref{SecIDM} and \ref{SecSPT}
introduce
the model and formulate the saddle-point approximation to it; its solution in
the limit of uniform droplet sizes is presented in Section \ref{SecUDL};
Sections \ref{Sec1IL}, \ref{SecIRFP} and \ref{SecFOT} are devoted to solution
of the model incorporating the broad distribution of droplet sizes; and
Section \ref{SecCD} is for summary and discussion.

\section{THE MODEL OF INTERACTING DROPLETS}
\label{SecIDM}

\begin{figure}[t]
{\centering{\includegraphics[scale=1.0]{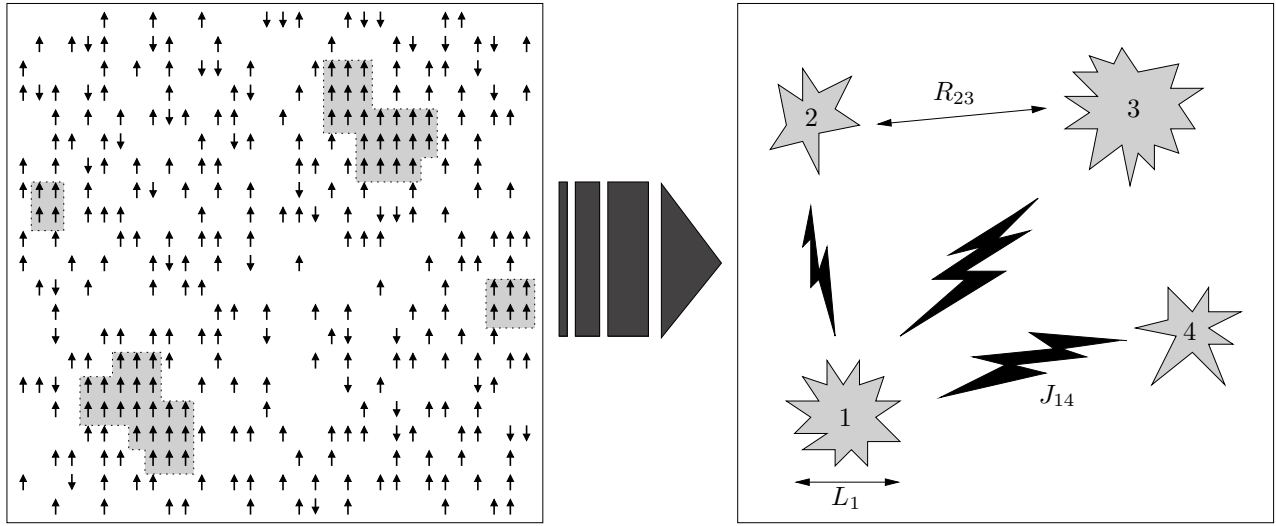}}\par}
\caption{Schematic depiction of the coarse-graining procedure leading to the
interacting droplet model of disordered magnetic systems.  Rare, defect-free
regions in the original system (shaded, right hand panel) are identified as
the important degrees of freedom.  Individual droplets then interact via
long-ranged interactions mediated by the metallic bulk.}
\label{GF}
\end{figure}

By focusing on locally ordered droplets as the relevant degrees of
freedom\cite{vojta:107202,vojta:045438},
a coarse-grained action in terms of a
fluctuating, local order parameter can be devised\cite{dobrosavljevic:187203}
to study the effect of
disorder on phase transitions in random magnets.  This coarse-graining
procedure is shown schematically in Figure \ref{GF}.  In this framework,
a single, isolated
droplet acts as a classical moment with dynamics in imaginary time and can be
mapped onto a classical, one-dimensional spin chain
with a local coupling
constant determined by the size of the droplet.
The local action for an isolated droplet is then given by
\begin{equation}
\label{Si}
S_{{\rm L},i}=\int_0^{\beta}{\rm
d}\tau{\rm d}\tau'\phi_i(\tau)\Gamma_i(\tau-\tau')\phi_i(\tau')
+\frac{u}{2N}\int_0^{\beta}{\rm d}\tau\phi_i^4(\tau)
\end{equation}
where $\phi_i$ is an $N$-component order parameter field for the $i$th droplet
and $\Gamma_i(\tau)$ is a generic coarse-grained two-point vertex.  For
definiteness, we will focus on the itinerant Heisenberg
antiferromagnet\cite{Hertz:1976kl} ($N>1$) so that
\begin{equation}
\label{GAF}
\Gamma_i(\omega_n)=(r_i+|\omega_n|);
\end{equation}
the $\omega_n$ are bosonic Matsubara frequencies $\omega_n=2n\pi T$ and the
local coupling constant is given by $r_i\approx \delta(L_i)^d$ where $L_i$ is
the
linear dimension of the droplet, $d$ is the spatial dimension, and $\delta<0$
is the coupling constant of the
clean magnet.  Droplet sizes follow a Poisson distribution
$P(L_i)\sim\exp[-\rho(L_i)^d]$ so that, correspondingly,
$P(r_i)\sim\exp[-\rho r_i/\delta]$, where $\rho$ is the volume fraction of
droplets.

In the absence of interactions between droplets, the action described by
(\ref{Si}) and (\ref{GAF}) is well understood\cite{vojta:045438}.
Ohmic dissipation given by
the linear $\omega_n$ term corresponds to long-ranged $1/\tau^2$ interactions
in imaginary time so that the corresponding Heisenberg spin chain is exactly
at its lower critical dimension\cite{Bruno:2001id} and the resulting energy
gap is found to be\cite{vojta:107202,vojta:045438}
$\epsilon_i\sim\exp[\pi r_i/u]$.  This leads to a power-law distribution of
local energies scales $P(\epsilon_i)\sim\epsilon_i^{\alpha-1}$ where the
Griffiths exponent $\alpha=d/z'$ is a non-universal function of parameters and
is expected to decrease as the magnetically ordered phase is approached.
Standard quantum Griffiths phase (QGP)
phenomenology follows from this distribution;
for example, the average susceptibility $\chi\sim T^{\alpha-1}$ diverges
as $T\rightarrow0$ for $\alpha<1$.

The above scenario applies in the absence of interactions between droplets
and, in particular, is valid for insulating magnets.  However, in itinerant
systems, electrons in the bulk mediate long-ranged interactions which must be
included.\cite{dobrosavljevic:187203}  Formally, the coarse-graining procedure
requires that the order
parameter fluctuations in the metallic bulk be integrated out, leading to an
effective RKKY interaction between droplets which has the asymptotic form
\begin{equation}
S_{{\rm I},ij}=\frac{J_{ij}}{(R_{ij})^d}\int_0^{\beta}{\rm d}\tau
\phi_i(\tau)\phi_j(\tau)
\end{equation}
where $R_{ij}$ is the distance between droplets $i$ and $j$, and $J_{ij}$ can
be taken to be a random amplitude with zero mean and variance
$\left<J_{ij}^2\right>=J^2$.
The effective description of a random itinerant magnet near a phase
transition is then given by the full action
\begin{equation}
\label{Sfull}
S=\sum_iS_{{\rm L},i}+\sum_{i,j}S_{{\rm I,}ij}.
\end{equation}

Recently,\cite{dobrosavljevic:187203} action (\ref{Sfull}) was studied for the
Heisenberg
antiferromagnet and it was found that infinitesimal interactions can
destabilize the QGP found in the non-interacting theory, leading to the
formation of the cluster glass phase (CGP).  These results showed that the
interactions can have a non-trivial effect on the droplet dynamics, even at
the saddle-point level which is formally justified in the large-$N$ limit.  In
the following Section, we review the saddle-point theory for the action
(\ref{Sfull}).

\section{SADDLE POINT THEORY}
\label{SecSPT}

By including RKKY interactions between separated droplets, individual droplets
no longer have only their own dynamics given by the kernel (\ref{GAF}), but
also acquire dynamics through their interactions with an effective bath given
by the collective fluctuations of all other droplet degrees of freedom.  To
obtain a description for this process, we average over the random
variables $J_{ij}$ using the standard method of introducing $n$ copies or
{\it replicas} of the system\cite{Binder:1986lm}
described by (\ref{Sfull})
and applying the identity
\begin{equation}
\left<\ln Z\right>_{\rm dis}=\lim_{n\rightarrow0}
\frac{\left<Z^n\right>_{\rm dis}-1}{n}
\end{equation}
where the partition function $Z={\rm Tr}\{\exp[-S]\}$ and
$\left<\cdots\right>_{\rm dis}$ represents averaging over random $J_{ij}$.  The
interaction term of (\ref{Sfull}) then becomes
\begin{equation}
\label{Sijeff}
S_{{\rm I},ij}=-\frac12\frac{J^2}{(R_{ij})^{2d}}\sum_{a,b=1}^n\int_0^{\beta}
{\rm d}\tau{\rm d}\tau'
\phi_i^a(\tau)\phi_i^b(\tau')\phi_j^a(\tau)\phi_j^b(\tau')
\end{equation}
which can be decoupled by introducing a Hubbard-Stratonovich
field\cite{Negele:1998du}
$Q^{ab}_i(\tau-\tau')$.  The replicated partition function can then be written
as
\begin{equation}
\left<Z^n\right>=\int{\rm D}\phi{\rm D}Q \exp\{-S_{\rm eff}[\phi,Q]\}
\end{equation}
with effective action
\begin{eqnarray}
\label{Seff}
S_{\rm
eff}[\phi,Q]&=&\sum_{a=1}^n\sum_{i,\omega_n}S_{{\rm L},i}[\phi_i^a]
+\frac12\sum_{a,b=1}^n\sum_{i,\omega_n}\phi_i^a(\omega_n)
\left[\sum_j\frac{J^2}{(R_{ij})^{2d}}Q_j^{ab}(\omega_n)\right]
\phi_i^b(\omega_n)\\ \nonumber
&&+~\frac12\sum_{a,b=1}^n\sum_{i,j,\omega_n}Q_i^{ab}(\omega_n)
\left(\frac{J^2}{(R_{ij})^{2d}}\right)Q_j^{ab}(\omega_n).
\end{eqnarray}
Defining the ``cavity field''
\begin{equation}
\Delta_i^{ab}(\omega_n)=\sum_j\frac{J^2}{(R_{ij})^{2d}}Q_j^{ab}(\omega_n),
\end{equation}
we see that the problem of interacting droplets has been reduced to a new
single-site problem, given by
\begin{equation}
S_{\rm L,
eff}[\phi,Q]=\sum_{a=1}^n\sum_{i,\omega_n}S_{{\rm L},i}[\phi_i^a]
+\frac12\sum_{a,b=1}^n\sum_{i,\omega_n}\phi_i^a(\omega_n)
\Delta_i^{ab}(\omega_n)
\phi_i^b(\omega_n),
\end{equation}
so that the effect of the other degrees of freedom is simply to
introduce additional dissipation through the frequency dependence of
$\Delta(\omega_n)$.  At the saddle-point level, the cavity field is determined
self-consistently from the relation
\begin{equation}
\label{Delta}
\Delta_i^{ab}(\omega_n)=\sum_j\frac{J^2}{(R_{ij})^{2d}}
\left<\phi_j^a(\omega_n)\phi_j^b(\omega_n)\right>_{\rm L}
\end{equation}
where the expectation value is taken with respect to $S_{\rm L, eff}$.
In the paramagnetic phase,
$\left<\phi_i^a(\omega_n)\phi_j^b(\omega_n)\right> = \delta_{ab}\chi_{\rm
L}(r_i,\omega_n)$ is diagonal in replica indices.  Noting also that the $r_i$
are independent of droplet position, (\ref{Delta}) is equivalent to
\begin{equation}
\Delta_i^{ab}(\omega_n)=\tilde{g}\delta_{ab}\overline{\chi_{\rm L}(\omega_n)}
=\tilde{g}\delta_{ab}\int{\rm d}r_iP(r_i)\chi_{\rm L}(r_i, \omega_n)
\end{equation}
where $\tilde{g}\equiv J^2\sum_{ij}(R_{ij})^{-2d}$.

To finalize the formulation of the saddle-point theory, the correlator
$\chi_{\rm L}$ can be calculated in the large-$N$ limit by decoupling the
quartic term in (\ref{Si}) with an auxiliary Hubbard Stratonovich field
$\lambda_i$.  The equations governing the dynamics of the droplets in the
interacting theory then become
\begin{eqnarray}
\label{SCX}
\overline{\chi_{\rm L}(\omega_n)}&=&\frac12\int{\rm d}r_i\frac{P(r_i)}
{r_i+\lambda_i+|\omega_n|-\tilde{g}\overline{\chi_L(\omega_n)}}\\
\label{SCl}
\lambda_i&=&\frac{uT}{2}\sum_{\omega_n}\frac{1}{r_i+\lambda_i+|\omega_n|
-\tilde{g}\overline{\chi_{\rm L}(\omega_n)}}
\end{eqnarray}
which must be evaluated self-consistently for $\lambda_i$ and
$\overline{\chi_{\rm L}(\omega_n)}$.

\section{QUANTUM CRITICALITY FOR UNIFORM DROPLETS} 
\label{SecUDL}

To understand the transition in the interacting droplet model, it is
instructive to first study equations (\ref{SCX}) and (\ref{SCl}) in the limit
of uniform droplet size, $P(r_i)=\delta(r_i-\hat{r})$.  In this limit, the
problem becomes equivalent to that of a quantum spin glass which 
is known to undergo a conventional second-order
transition from a paramagnetic Fermi liquid (PMFL) to a spin glass at large
enough $\tilde{g}$.\cite{sachdev}

Self-consistency equation (\ref{SCX}) reduces to an algebraic
equation for $\chi(\omega_n)$ and is easily solved:
\begin{equation}
\label{ChiU}
\chi_{\rm U}(\omega_n)=\frac1{2\tilde{g}}\left(
\hat{r}+\hat{\lambda}+|\omega_n|\pm\sqrt{(\hat{r}+\hat{\lambda}+|\omega_n|)^2
-2\tilde{g}}\right)
\end{equation}
where $\hat{\lambda}\equiv\lambda(\hat{r})$.  Clearly, this solution becomes
unstable when the energy scale defined by
$\Delta\equiv(\hat{r}+\hat{\lambda}-\sqrt{2\tilde{g}})$ becomes zero.  The
phase boundary can then be determined by setting $\Delta=0$ and solving
(\ref{SCl}) self-consistently for $\hat{r}_c(T)$ (or, equivalently,
$\hat{\lambda}_c(T)$).  
In particular, at $T=0$, this gives
\begin{equation}
\hat{r}_c(T=0)=\sqrt{2\tilde{g}}+\frac{u}{2\pi}\left(
\frac12+\ln\left(\frac{\sqrt{2\tilde{g}}}{2\Lambda}\right)\right)
\end{equation}
where $\Lambda$ is an ultra-violet cutoff.  In the limit
$\tilde{g}\rightarrow0$, $\hat{r}_c(0)\rightarrow-\infty$ showing that
frustration induced by the interactions is necessary to stabilize the glass
phase.

The phase for $\Delta>0$ ($\hat{r}>\hat{r}_c$) is characterized by the
frequency dependence of
$\chi_{\rm U}$.  From the solution (\ref{ChiU}), we find three separate
regimes:
\begin{equation}
\chi_{\rm U}(0)-\chi_{\rm U}(\omega)\sim
\left\{
\begin{array}{cc}
\omega&,\omega<\Delta\ll\sqrt{2\tilde{g}}\\
\sqrt{\omega}&,\Delta<\omega\ll\sqrt{2\tilde{g}}\\
\chi_{\rm U}(0)-\omega^{-1}&,\omega\gg\sqrt{2\tilde{g}}
\end{array}
\right..
\end{equation}
The linear, low-$\omega$ dependence of $\chi_{\rm U}$ for $\Delta>0$ is
characteristic of a
Fermi liquid, while, right at the transition, $\chi_{\rm U}$ behaves as
$\sqrt{\omega}$.  This sub-linear form at $\Delta=0$ leads to
non-Fermi liquid behavior of all thermodynamic quantities, though these
violations are fairly mild since the corrections remain finite at $T=0$.
Thus, there is no quantum Griffiths behavior in this model since we have
neglected the fluctuations arising from a distribution of droplet sizes.

The transition at $T=0$ can also be approached from the magnetically
ordered side.
In this case, the quantum critical point can
be determined when the mean-field stability
criterion for the glass phase vanishes\cite{pastor:4642}, i.e.:
\begin{equation}
\label{lamsg1}
\lambda_{\rm
SG}\equiv1-\sqrt{2\tilde{g}}\chi_{\rm U}(0)=0.
\end{equation}
Using the solution (\ref{ChiU}) for $\chi_{\rm U}$, we find $\lambda_{\rm
SG}\sim\sqrt{\Delta}$ which vanishes at $\Delta=0$.

Thus, examining the transition through the instability of either the
paramagnetic
Fermi liquid or the spin-glass phase yields a quantum critical point at
$\Delta=0$, consistent with a conventional second-order transition in this
problem.  In the next Sections, we will consider the model with the full
distribution of droplet sizes and show that the two instability criteria do
not coincide, revealing the singular effect of droplet fluctuations.

\section{DISTRIBUTED DROPLETS AND THE CLUSTER GLASS PHASE}
\label{Sec1IL}

For dilute impurities, the defect-free regions assume a Poisson distribution
which can be written in terms of the local coupling constant as
\begin{equation}
P(r_i)=\frac{2\pi\kappa}{u}e^{2\pi\kappa(r_i-\hat{r})/u},\hspace{1em}
r_i\le\hat{r}<0.
\end{equation}
The offset $\hat{r}$ is used to tune through the transition and $\kappa=\rho
u/2\pi\delta$.  The uniform limit of the previous Section is recovered for
$\kappa\rightarrow\infty$; however, by accounting for droplets of all sizes, we
are also including their fluctuations which are associated with Griffiths
phase behavior.

Using the uniform solution (\ref{ChiU}) as a zeroth order approximation, we
can iterate equations (\ref{SCX}) and (\ref{SCl}) until self-consistency is
achieved.  Already at one iteration loop (1IL), the instability of the
Griffiths phase at $T=0$ is apparent.  At this level of approximation,
equation (\ref{SCl}) can be integrated exactly.  Identifying the local droplet
energy $\epsilon_i=\hat{r}+r_i+\lambda_i-\tilde{g}\chi_{\rm U}(0)$, we find
the relation between energy scale and local coupling strength,
$\epsilon_i\propto\exp[2\pi f^{-1}r_i/u]$ where
\begin{equation}
\label{fff}
f\equiv\frac{2\sqrt{\Delta^2+2\Delta\sqrt{2\tilde{g}}}}
{\Delta+\sqrt{2\tilde{g}}+\sqrt{\Delta^2+2\Delta\sqrt{2\tilde{g}}}}.
\end{equation}
This allows us to switch from integration over $r_i$ in equation (\ref{SCX}),
to integration over $\epsilon_i$ via the replacement $P(r_i){\rm
d}r_i\rightarrow \widetilde{P}(\epsilon_i){\rm d}\epsilon_i$ where
\begin{equation}
\label{epsdist}
\widetilde{P}(\epsilon_i)=\left(\kappa fe^{-2\pi\kappa\hat{r}/u}\right)
\epsilon^{\kappa f-1}.
\end{equation}
Defining $\alpha\equiv\kappa f$, (\ref{epsdist}) takes the power-law form
characteristic of droplets in the Griffiths phase with Griffiths exponent
$\alpha$.  The novelty arises when this distribution is used to calculate the
frequency dependent cavity field,
\begin{eqnarray}
\overline{\chi_{\rm L}
(\omega_n)}&=&\frac{\alpha}{2}e^{-2\pi\kappa\hat{r}/u}\int_0^{\Lambda_\epsilon}
{\rm
d}\epsilon_i\frac{\epsilon_i^{\alpha-1}}{\epsilon_i+|\omega|-
\tilde{g}(\chi_{\rm
U}(\omega)-\chi_{\rm U}(0))},
\end{eqnarray}
where $\Lambda_{\epsilon}$ is an upper cutoff in energy.
This gives
\begin{equation}
\overline{\chi_{\rm L}(\omega)}-\overline{\chi_{\rm L}(0)}=
-\gamma|\omega|^{\alpha-1}+\cal{O}(|\omega|),
\end{equation}
where $\gamma$ is a constant.
Thus, droplets for which $\alpha<2$ acquire non-Ohmic dissipation and the
corresponding Heisenberg spin chains find themselves above their lower
critical dimension\cite{Bruno:2001id}
and can order in the imaginary time direction.  Due to the
frustration caused by the effectively random RKKY interaction between these
ordered droplets, they form a glassy state dubbed the "cluster glass phase"
(CGP).\cite{dobrosavljevic:187203}
Importantly, the CGP is present over the entire range where one
would predict the QGP from the isolated droplet model.\cite{vojta:045438}

\begin{figure}[t]
{\centering{\includegraphics[scale=0.75]{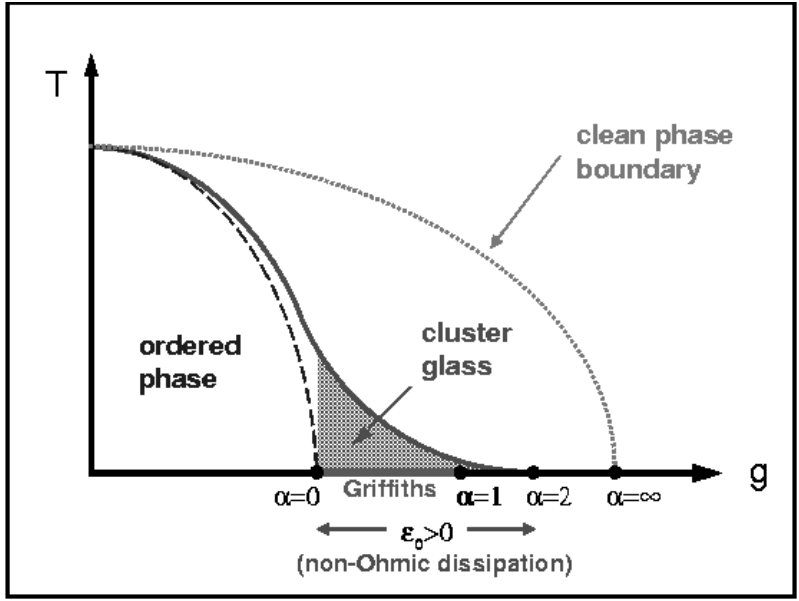}}\par}
\caption{Schematic phase diagram of random itinerant magnets.  Interactions
destabilize the putative quantum Griffiths phase at $\alpha=1$, favoring the
formation of the cluster glass phase for $\alpha<2$.  The phase boundary has
an exponential tail due to the exponentially rare nature of the large frozen
droplets.}
\label{VPD}
\end{figure}

The finite temperature phase boundary can also be estimated in this formalism.
The critical coupling constant is determined from (\ref{SCl}) at
$\epsilon_i=0$:
\begin{eqnarray}
\label{ric}
r_{i,\rm c}&=&-\frac{u}{\pi}\int_0^{\Lambda}{\rm
d}\omega\frac{1}{\omega+\gamma\tilde{g}\omega^{\alpha-1}}\\ \nonumber
&\approx&-\frac{u}{\pi(2-\alpha)}\ln(1/\gamma\tilde{g})
\end{eqnarray}
and the number of frozen droplets is $n_{\rm fr}\sim\exp(-\rho
r_{i, \rm c}/\delta)$.  The critical temperature is then estimated
as\cite{vojta:107202}
\begin{equation}
T_{\rm c}\sim n_{\rm fr}\sim\exp\left[-\frac{\rho u}{\pi|\delta|(2-\alpha)}
\log(1/\gamma\tilde{g})\right]
\end{equation}
which has an exponential tail as $\alpha\rightarrow2$.  The phase diagram is
shown schematically in Figure \ref{VPD}.

\section{IRFP OF ISOLATED DROPLETS AT QUANTUM CRITICALITY}
\label{SecIRFP}

It is interesting to note that our 1IL expression for the Griffiths exponent
hints at the close analogy between the present work and the physics of
infinite-randomness fixed points (IRFP).\cite{Fisher:1992qf}
From the expression
$\alpha=f\kappa$ with $f$ given in equation (\ref{fff}), we would conclude at
this level of approximation that the Griffiths exponent vanishes at the uniform
droplet quantum critical point as
\begin{equation}
\alpha\sim\sqrt{\Delta}.
\end{equation}
This is significant in that it gives a simple access to the IRFP that has been
lacking in the past, while capturing the physics of the quantum Griffiths
phase.  Of course, this expression for $\alpha$ is only valid at 1IL, and, in
particular, does not hold at the lowest temperatures where the system freezes
into the CGP.  However, QGP and IRFP behavior can still be expected at
temperatures above the CGP transition temperature.  We will address this issue
further in the final Section.

\section{FLUCTUATION INDUCED FIRST ORDER TRANSITION} 
\label{SecFOT}

In the paramagnetic phase with $\alpha>2$, the leading behavior of $\chi_{\rm
L}(\omega)$ is linear in $\omega$.  Writing $\overline{\chi_{\rm L}(\omega)}
-\overline{\chi_{\rm
L}(0)}=-m_{\chi}\omega$ and inserting this into the self-consistency condition
(\ref{SCl}) gives
\begin{eqnarray}
\epsilon_i-r_i-\hat{r}+\tilde{g}\overline{\chi_{\rm L}(0)}&=&\frac{u}{2\pi}
\int_0^{\Lambda}{\rm d}\omega\frac1{\epsilon_i+(1-m_{\chi})|\omega|}\\
\nonumber
&=&\frac{u}{2\pi(1-\tilde{g}m_{\chi})}
\ln\left(\frac{\epsilon_i+(1-m_{\chi})\Lambda}{\epsilon_i}\right).
\end{eqnarray}
Analyzing this for $r_i\ll0$ gives
\begin{equation}
\epsilon_i\sim \exp\left(\frac{2\pi(1-\tilde{g}m_{\chi})}{u}\cdot r_i\right)
\end{equation}
and, comparing to the 1IL results, defines
\begin{eqnarray}
\label{ffull}
f&\equiv&(1-\tilde{g}m_{\chi})^{-1};\\
\label{alphdef}
\alpha&=&f\kappa,
\end{eqnarray}
which can be checked
against the previous results.  Note that expression (\ref{alphdef}) is exact
in the paramagnetic phase provided that $m_{\chi}=\partial\overline{\chi_{\rm
L}(\omega)}/\partial\omega|_{\omega=0}$ is determined self-consistently, so,
following the arguments of Section \ref{Sec1IL},
the instability of the PMFL to CGP formation still corresponds to $\alpha=2$,
to all orders.

This relation can be further illuminated by calculating the slope exactly from
(\ref{SCX}):
\begin{eqnarray}
m_{\chi}&=&-\frac12\int{\rm
d}r_i\frac{(1-\tilde{g}m_{\chi})P(r_i)}
{[r_i+\lambda_i-\tilde{g}\chi_{\rm L}(0)]^2}\\ \nonumber
&=&-2(1-\tilde{g}m_{\chi})\overline{\chi^2_{\rm L}(0)}.
\end{eqnarray}
Solving this for $m_{\chi}$, we get
\begin{equation}
\label{lamsg2}
1-\tilde{g}m_{\chi}=(1-2\tilde{g}\overline{\chi^2_{\rm L}(0)})^{-1}.
\end{equation}
What is striking about this result is that the term in parentheses on the
right hand side of (\ref{lamsg2}) is the exact analogue of the stability
criterion (\ref{lamsg1}) for droplets with distributed site energies;
\cite{pastor:4642} in this
case, $\lambda_{\rm SG}=1-2\tilde{g}\overline{\chi_{\rm L}^2(0)}$.  Relating
this to the definition (\ref{alphdef}) for $\alpha$, we can write
\begin{equation}
\alpha=\lambda_{\rm SG}\kappa
\end{equation}
so that $\lambda_{\rm SG}>0$ at the point where the PMFL becomes unstable
($\alpha=2$).  This should be contrasted with the uniform droplet limit of
Section \ref{SecUDL} where the instability of the PMFL and the instability
of the spin-glass phase coincided at a conventional second-order transition.
In the present case, the stability criteria signify
spinodal lines and the CGP and
PMFL coexist in a region around a first-order phase transition.

To gain more insight into the nature of the transition, we recast the problem
as an eigenvalue analysis of the constitutive free energy for which the
solution to (\ref{SCX}) is a minimum.  The relevant contribution to the
free energy can be expressed near the minimum as
\begin{equation}
{\cal F}_{\chi}=\int{\rm d}\omega{\rm d}\omega'(\chi_{\rm L}(\omega)-
\chi_0(\omega))\Gamma_{\chi}(\omega,\omega')(\chi_{\rm L}(\omega')-
\chi_0(\omega'))
\end{equation}
so that the eigenvalue determining the stability of the PMFL solution $\chi_0$
satisfies
\begin{equation}
\lambda_{\chi}\le
\frac{\int{\rm d}\omega{\rm d}\omega'(\chi_{\rm L}(\omega)-
\chi_0(\omega))\Gamma_{\chi}(\omega,\omega')(\chi_{\rm L}(\omega')-
\chi_0(\omega'))}{\int{\rm d}\omega(\chi_{\rm L}(\omega)-\chi_0(\omega))^2}.
\end{equation}
The self-consistency condition (\ref{SCX}) for $\chi$ results from the
functional
derivative $\delta{\cal F}_{\chi}/\delta\chi=0$ so that
\begin{equation}
\int{\rm d}\omega'\Gamma_{\chi}(\omega,\omega')\chi_{\rm
L}(\omega')=\chi_{\rm L}(\omega)-\int{\rm d}r_i
\frac{P(r_i)}{r_i+\lambda_i+|\omega|-\tilde{g}\chi_{\rm L}(\omega)}.
\end{equation}
and 
\begin{equation}
\int{\rm d}\omega'\Gamma_{\chi}(\omega,\omega')\chi_0
(\omega')=0.
\end{equation}
Now, define an eigenvalue $\lambda^{(n)}$ valid at each step in an iteration
scheme
\begin{eqnarray}
\lambda_{\chi}^{(n)}&\le&\frac{\int{\rm d}\omega{\rm d}\omega'(\chi_{\rm
L}^{(n-1)}(\omega)-\chi_{\rm L}^{(n)}(\omega))
\Gamma_{\chi}(\omega,\omega')(\chi_{\rm L}^{(n-1)}(\omega')
-\chi_{\rm L}^{(n)}(\omega'))}{\int{\rm d}\omega(\chi_{\rm
L}^{(n-1)}-\chi_{\rm L}^{(n)}(\omega))^2}\\ \nonumber
&=&1-\frac{\int{\rm d}\omega(\chi_{\rm L}^{(n-1)}-\chi_{\rm L}^{(n)}(\omega))
(\chi_{\rm L}^{(n)}(\omega)-\chi_{\rm L}^{(n+1)}(\omega))}
{\int{\rm d}\omega(\chi_{\rm L}^{(n-1)}(\omega)-\chi_{\rm L}^{(n)}(\omega))^2}
\end{eqnarray}
and proceed numerically.  From this, the spinodal line defined by
$\lambda_{\chi}=0$ can be determined at zero and finite temperatures, and
compared with the condition $\lambda_{\rm SG}=0$.

\begin{figure}[t]
{\centering{\includegraphics[scale=1.0]{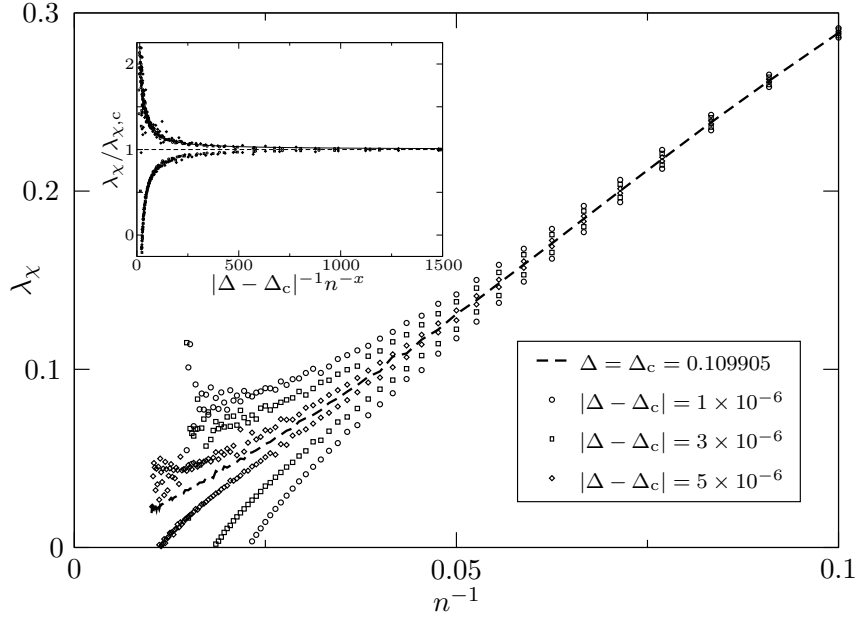}}\par}
\caption{Plot of the stability criterion $\lambda_{\chi}$ versus
$n^{-1}$ for several values of $\Delta$.  The dashed line is the
curve at the critical $\Delta_{\rm c}$.  Data above this line are
for $\Delta>\Delta_{\rm c}$ while data below are for
$\Delta<\Delta_{\rm c}$. The inset demonstrates scaling of the
curves, as discussed in the text.} \label{lamplots}
\end{figure}

The results of this procedure at $T=0$
are shown in Figure (\ref{lamplots}).  As the transition is
approached from the PMFL, $\lambda_{\chi}$ appears to approach a positive
value as $n\rightarrow\infty$.  We identify the critical value of $\Delta_{\rm
c}$
at the transition as the last point where the iterative method converges; the
associated curve $\lambda_{\chi, \rm c}~vs.~n^{-1}$ is indicated by the dashed
line in the figure and extrapolates to zero at $n^{-1}=0$.  Below the
transition, $\lambda_{\chi}$ eventually becomes negative at some finite
iteration step and the procedure breaks down.  The inset to Figure
(\ref{lamplots}) shows a scaling analysis for the same data as in the main
panel.  Using the {\it ansatz} $\lambda_\chi/\lambda_{\chi,\rm
c}=\Lambda_{\pm}(|\Delta-\Delta_{\rm c}|n^x)$, we find excellent scaling with
exponent $x=2.35$.  While the meaning of the exponent is unclear at this
point, the scaling shown in Figure (\ref{lamplots}) demonstrates that the
iterative method is well controlled.

\begin{figure}[t]
{\centering{\includegraphics[scale=1.0]{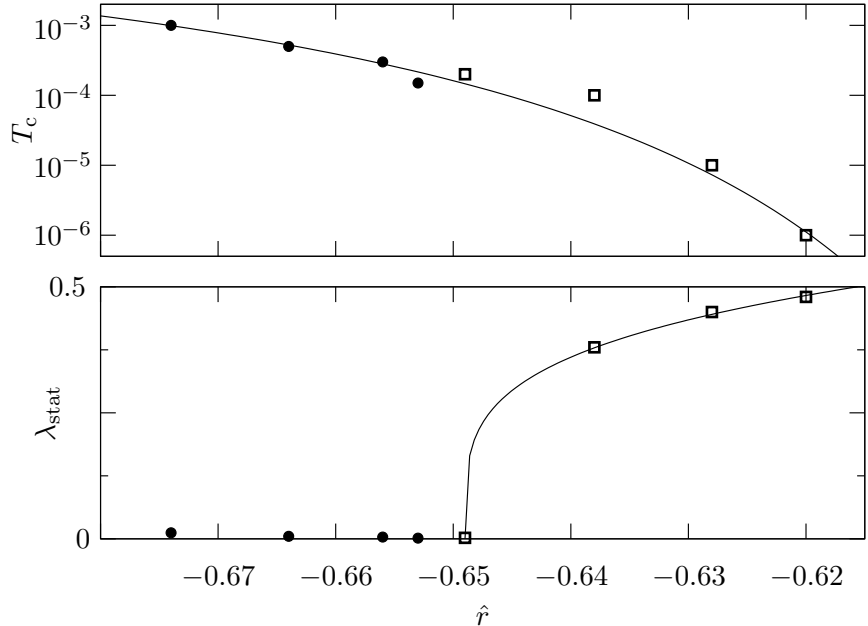}}\par}
\caption{{\it Upper panel}: The phase boundary at finite
temperatures is shown by the symbols and corresponds to
$\lambda_{\chi}=0$.  The line is a fit to the data as discussed in
the text. {\it Lower panel}: The static stability criterion
calculated at $T_{\rm c}$.  The line is a guide to the eye.  Empty
squares correspond to first-order transitions, full circles to
second-order transitions.}
\label{Tricrit}
\end{figure}

At finite temperatures, we find that the two stability criteria continue to
disagree over a portion of the phase boundary away from $T=0$, as shown in
Figure (\ref{Tricrit}).  The upper panel shows the boundary line defined by
$\lambda_{\chi}=0$.  The solid line is a fit to the data using the form
$\log(T)\sim(\hat{r}-\hat{r_{\rm c}})^{-1}$, reflecting the fact that the
critical temperature is proportional to the number of frozen droplets, as
discussed in Section (\ref{Sec1IL}).  The value of $\lambda_{\rm SG}$ at the
boundary defined by $\lambda_{\chi}=0$ is shown in the lower panel.  We can
see that $\lambda_{\rm SG}>\lambda_{\chi}=0$ for
$-0.65\le\hat{r}\le\hat{r}_{\rm c}(0)$.  For $\hat{r}<-0.65$, the two criteria
merge, indicating that the standard continuous transition is restored.  In
both panels, empty squares represent the first-order phase boundary and full
circles represent the second-order phase boundary.

This scenario is consistent with a first-order transition induced by the
fluctuations associated with the Griffiths phase.  In the uniform
limit, these fluctuations were suppressed since all droplets were of equal
size, and the transition was of a conventional, second-order type.  However,
with the full distribution of droplet sizes included, the transition at $T=0$
was driven first-order.  At finite temperature, $1/T$ acts an effective cutoff
in the imaginary time direction, suppressing fluctuations of the largest
droplets.  Thus, the restoration of the second-order transition at high enough
temperatures is a natural consequence of eliminating fluctuations associated
with the longest time-scales in the exponential tail of the distribution, and
we conclude that these fluctuations are responsible for the discontinuous
nature of the transition at $T=0$ and above.

\begin{figure}[t]
{\centering{\includegraphics[scale=1.0]{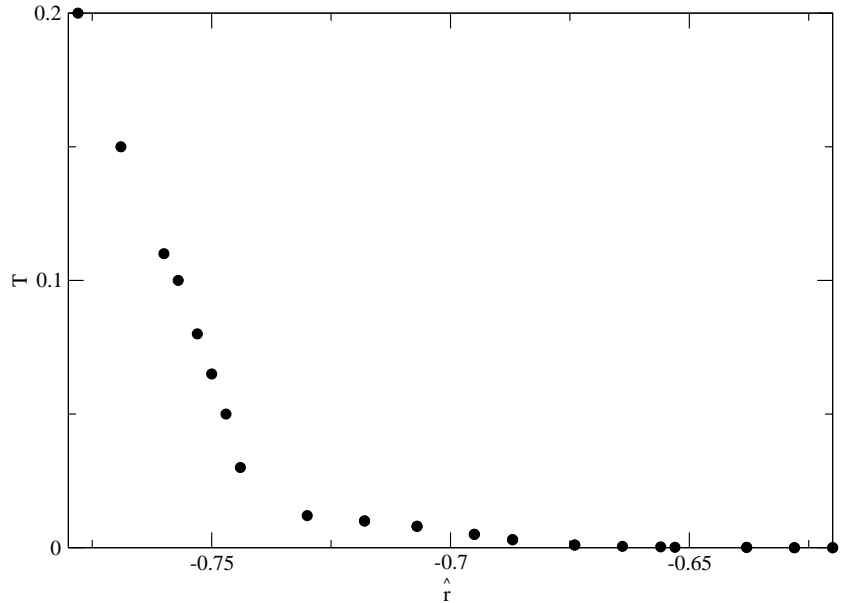}}\par}
\caption{The numerically determined phase boundary on a linear temperature
scale showing the exponential tail due to freezing of rare large droplets.}
\label{PB}
\end{figure}

Finally, in Figure (\ref{PB}), we show the phase boundary on a linear
temperature scale and for higher temperatures.  At low temperatures, there is
an exponential tail reflecting the fact that the largest droplets which freeze
to form the CGP are exponentially rare in this regime, as was shown
schematically in Figure (\ref{VPD}).

\section{CONCLUSIONS AND DISCUSSION}
\label{SecCD}

The fluctuations of rare, defect-free regions in random magnets can lead to a
variety of interesting phenomena near quantum critical points.  In the present
work, we focused on itinerant antiferromagnets with a $T=0$ transition
between the PMFL and CGP, and saw that the Griffiths anomalies provide a
novel mechanism responsible for driving the transition first-order.

This work could be of relevance to experiments on random systems in metallic
environments.  For example, it has recently been shown\cite{Vafek:2006fk}
that singular
corrections due to rare regions are necessary in understanding the transition
in a magnetic field of superconducting thin films in the Little-Parks
experiment at $T=0$.\cite{Liu:2001lr}
The non-Fermi liquid behavior arising due to quantum Griffiths anomalies may
also be relevant for certain strongly correlated materials.
Recent studies on
such varied systems as the rare-earth intermetallic
Tb$_5$Si$_2$Ge$_2$\cite{magen:167201} and the
colossal magnetoresistive
La$_{0.7}$Ca$_{0.3}$MnO$_3$\cite{PhysRevLett.88.197203} show power-law
diverging susceptibilities reminiscent of quantum Griffiths physics.
We conclude with a brief discussion of the temperature scales relevant to 
experimental observations of cluster glass physics.

From equation (\ref{ric}), we notice that the effects of non-Ohmic dissipation
are only apparent below a crossover frequency
\begin{equation}
\omega^*=\exp[-\ln(1/\gamma\tilde{g})/(2-\alpha)].
\end{equation}
Thus, for temperatures below this energy scale, thermodynamic quantities
should be dominated by the contribution due to frozen droplets, giving, for
example, $\chi\sim T^{-1}$.  Above this temperature scale, Ohmic dissipation
dominates and the results from the non-interacting droplet model will apply.
In particular, thermodynamics quantities will diverge as a power-law
with exponent $\alpha-1<0$ as in the quantum Griffiths phase.

\section*{ACKNOWLEDGMENTS}

We would like to acknowledge fruitful discussions with Andrey 
Chubukov, Eduardo Miranda, Joerg Schmalian, Oskar Vafek, and
Thomas Vojta. This work was supported through the NSF Grant No.
DMR-0542026 (V. D.) and the National High Magnetic Field
Laboratory (M.J.C.). We also thank the Aspen Center for Physics,
where part of this work was carried out and the School of Human Flight.


\end{document}